\documentclass[prl,twocolumn,showpacs,preprintnumbers,amsmath,amssymb]{revtex4}
\usepackage{bm, psfrag, graphics}
\begin{document}
\title{Symmetry-breaking self-consistent quantum many-body structure of high-lying macroscopic self-trapped and superposition states of the gaseous double-well BEC}
\author{David J. Masiello}
\email{masiello@u.washington.edu}
\author{William P. Reinhardt}
\email{rein@chem.washington.edu}
\affiliation{Department of Chemistry, University of Washington, Seattle, Washington 98195-1700, USA}
\date{\today}       
\begin{abstract}
The many-body structure of high-lying excited states, including macroscopic quantum superpositions, of the gaseous double-well BEC is presented within the context of a multiconfigurational bosonic self-consistent field theory based upon underlying symmetry-broken one-body wave functions. To better understand our initial results, a model is constructed in the extreme Fock state limit, in which macroscopic quantum self-trapped and superposition states emerge in the many-body spectrum, striking a delicate balance between the degree of symmetry breaking, the effects of the condensate's mean field, and that of atomic correlation. It is found that, in general, the superposition state lies energetically below its related self-trapped counterpart. Furthermore, noticeably different spatial density profiles are associated with each type of excited state.
\end{abstract}
\maketitle

Bose-Einstein condensates (BECs) of identical atoms confined to external double-well trapping potentials are actively being investigated \cite{Shin2004a,Shin2004b,Saba05,Albiez04}. Remarkably sensitive matter-wave interferometry \cite{Shin2004a}, based upon a coherently split gaseous BEC source, represents just one potentially technologically useful experimental tool already demonstrated in the double-well condensate. Recent experiments have also achieved exotic macroscopic quantum self-locked (self-trapped) states \cite{Albiez04} and, in doing so, open the way for direct experimental study of the fundamental nature of macroscopic quantum superpositions \cite{Leggett05} directly observed in the gaseous BEC while only more indirectly in more conventional areas of condensed matter experiment \cite{Friedman00,Lloyd00}. None of the above experiments have been modeled in any detail at a first principles level.

Theoretical descriptions of the double-well condensate's fully correlated many-body ground and low-lying excited states have been explored within the restricted Fock state basis $|N_1,N_2\rangle=(\hat b_1^\dagger)^{N_1}(\hat b_2^\dagger)^{N_2}|{\textrm{vac}}\rangle/\sqrt{N_1!N_2!},$ where $N_k$ is restricted to two macroscopically occupied states, $k=1,2.$ Here, the one-body wave functions (orbitals) $\chi_k$ approximating the bosonic field operators as $\hat\Psi({\bf x})=\chi_1({\bf x})\hat{b}_1+\chi_2({\bf x})\hat{b}_2$ have been chosen, with varying levels of approximation, from solutions of the one-body Schr\"odinger equation \cite{Spekkens1999a}, to solutions of the bosonic Hartree-Fock equations \cite{Masiello05}, where the necessity of inclusion of orbital mean-field effects was established, as well as with solutions that have reached self-consistency in the occupation numbers $n^\nu_k$ of the $\nu$th quantum state \cite{Ced06}. The roots of each of these approaches find their origin in the quantum chemistry of many-electron atomic and molecular systems \cite{Ostlund}.

It is the purpose of this Letter to provide a novel theoretical first principles modeling of these high-lying excited states of the gaseous double-well BEC within the framework of an extended multiconfigurational bosonic self-consistent field (MCBSCF) theory that allows for symmetry breaking in both the many-body wave function and in the underlying orbitals $\chi_k,$ and includes a complete and consistent description of the effects of the condensate's mean field and atomic correlation in the restricted Fock space. In this theory, solutions of the many-body Schr\"odinger equation are constructed in which the orbitals underlying the $N$-boson Hamiltonian $\hat{H}=\int\hat{\Psi}^\dagger({\bf x})h({\bf x})\hat{\Psi}({\bf x})d^{3}x+(1/2)\int\hat{\Psi}^\dagger({\bf x})\hat{\Psi}^\dagger({\bf x}')V({\bf x},{\bf x}')\hat{\Psi}({\bf x}')\hat{\Psi}({\bf x})d^{3}xd^{3}x',$ have reached self-consistency with the many-body density. Within our restricted Fock space, the most general state of the system is a superposition of multiple configurations of identical bosonic atoms distributed, in all possible ways, between two condensate states at zero temperature. It is of the form
\begin{equation}
\label{state}
|\Psi^N;\{{\bf C}^\nu;\alpha\}\rangle=\sum_{N_1=0}^{N}C^{\nu}_{N_1}[{\bf C}^\nu;\alpha]|N_1,N_2;\{{\bf C}^\nu;\alpha\}\rangle,
\end{equation}
where the nonlinear dependence of the orbitals (and the many-body state itself) upon the $\nu$th eigenvector of the Hamiltonian ${\bf C}^\nu=\{C_{N_1}^\nu:N_1=0,\ldots,N\}$ and the orbital symmetry-breaking parameter $\alpha$ are indicated in brackets. The novelty of this ansatz is that it extends the work of Ref. \cite{Ced06} to variationally include the effects of spatial symmetry-breaking, through $\alpha,$ in both the underlying orbital basis and in the macroscopic many-body wave function. The particular Fock basis state $|N_1,N_2;\{{\bf C}^\nu,\alpha\}\rangle$ may represent, {\it e.g.}, a symmetry-broken configuration with $N_1$ atoms in the left well and $N_2$ atoms in the right well of the double-well trapping potential or perhaps a configuration with $N_1$ and $N_2$ atoms in a symmetric and antisymmetric pair of states that enjoy the full symmetry of the Hamiltonian \cite{fn6}. The specific nature of this basis and its connection to $\alpha$ will be clarified in the following. For a particular excited state $\nu=0,\ldots,N,$ self-consistency is measured through the occupation numbers $n^\nu_k,$ which are the eigenvalues of the one-body reduced density matrix $\gamma^\nu({\bf x},{\bf x}'),$ {\it i.e.}, $\int\gamma^\nu({\bf x},{\bf x}')\chi_k({\bf x}')d^3x'=n^\nu_k\chi_k({\bf x}).$ Both the number of condensed atoms $N_k$ in and the occupation numbers $n^\nu_k$ of the $k$th Fock state sum to the total number of atoms $N.$ By solving the coupled MCBSCF orbital equations
\begin{widetext}
\begin{equation}
\begin{split}
\label{orb1}
\mu_{kk}\chi_k({\bf x})\gamma^\nu_{kk}+\mu_{kl}\chi_l({\bf x})\gamma^\nu_{kl}&=[h({\bf x})\gamma_{kk}^\nu+{\cal V}_{kk}({\bf x})\Gamma_{kkkk}^\nu+{\cal V}_{ll}({\bf x})\Gamma_{klkl}^\nu+{\cal V}_{kl}({\bf x})\Gamma_{kkkl}^\nu+{\cal V}_{lk}({\bf x})\Gamma_{lkkk}^\nu]\chi_k({\bf x})\\
&\ \ \ +[h({\bf x})\gamma_{kl}^\nu+{\cal V}_{kk}({\bf x})\Gamma_{kkkl}^\nu+{\cal V}_{ll}({\bf x})\Gamma_{klll}^\nu+{\cal V}_{kl}({\bf x})\Gamma_{kkll}^\nu+{\cal V}_{lk}({\bf x})\Gamma_{kllk}^\nu]\chi_l({\bf x})
\end{split}
\end{equation}
\end{widetext}
for $\chi_k,$ $k\neq l=1,2,$ and then using these orbitals to construct the many-body Hamiltonian $\hat H,$ subsequent full diagonalization of $\hat H$ in (\ref{state}) yields many-body solutions $|\Psi^N;\{{\bf C}^\nu,\alpha\}\rangle$ that may, by iteration, reach self-consistency in occupation number with the orbitals underlying the $\nu$th quantum state. In Eq. (\ref{orb1}), $h({\bf x})=(-\hbar^2/2m)\nabla^2+V_{\textrm{ext}}({\bf x})$ is the one-body Hamiltonian, ${\cal V}_{kl}({\bf x})=\int\chi_k^*({\bf x}')V({\bf x},{\bf x}')\chi_l({\bf x}')d^3x'$ are one-body matrix elements of the atom-atom interaction potential $V({\bf x},{\bf x}')=(4\pi\hbar^2a/m)\delta({\bf x}-{\bf x}')$ in the contact approximation \cite{Huang1987a} with repulsive $s$-wave scattering length $a,$ and $\gamma^\nu_{kl}$ and $\Gamma^\nu_{klmn}$ ($k,l,m,n=1,2$) are Fock space matrix elements of the one- and two-body reduced density matrices $\gamma^\nu({\bf x},{\bf x}')=\langle\hat\Psi^\dagger({\bf x})\hat\Psi({\bf x}')\rangle_\nu$ and $\Gamma^\nu({\bf x},{\bf x}';{\bf y},{\bf y}')=\langle\hat\Psi^\dagger({\bf x})\hat\Psi^\dagger({\bf x}')\hat\Psi({\bf y})\hat\Psi({\bf y}')\rangle_\nu$ \cite{Lowdin55a,Penrose56} taken with respect to the MCBSCF state (\ref{state}). These matrix elements are functions of the expansion coefficients ${\bf C}^\nu$ appearing in Eq. (\ref{state}) and, thus, are not specified but rather are connected to $\hat H$ through its eigenvectors. The field operators $\hat\Psi({\bf x})$ and $\hat\Psi^\dagger({\bf x}')$ and basic creation and annihilation operators $\hat b_k$ and $\hat b_l^\dagger$ satisfy the usual bosonic commutation relations.


Before discussing the MCBSCF theory of macroscopic quantum self-trapped and superposition states of the BEC, it is instructive to consider a simplified model within multiconfigurational bosonic Hartree-Fock (MCBHF) theory \cite{Masiello05}. By omitting the off-diagonal couplings in only the orbital sector of the MCBSCF theory, solutions of the resulting bosonic Hartree-Fock (BHF) equations are used to construct the many-body Hamiltonian. Subsequent full diagonalization of ${\hat H}$ in (\ref{state}) yields solutions that may, similarly, achieve self-consistency through the diagonal parts of the density, or through application of the Hylleraas-Undheim theorem \cite{Hyll30}. It is the latter approach that we employ in the following model while we strive for self-consistency in the general theory.

{\it Approximate implementation of restricted Fock space model.} The diagonal terms in Eq. (\ref{orb1}) form the generalized BHF equations
\begin{equation}
\label{GHF}
[h\gamma_{kk}^\nu+{\cal V}_{kk}\Gamma_{kkkk}^\nu+{\cal V}_{ll}\Gamma_{klkl}^\nu]\chi_k+{\cal V}_{lk}\Gamma_{kllk}^\nu\chi_l=\mu_{kk}\chi_k\gamma^\nu_{kk}
\end{equation}
for $N$ identical bosons where, again, $k\neq l=1,2.$ We point out that the matrix elements of $\gamma^\nu$ and $\Gamma^\nu$ appearing in Eq. (\ref{GHF}) are not specified by BHF mean-field theory, but rather find their origin in the full many-body theory where the state of the system may, in general, be a linear combination of several configurations. This fact will allow us, in the next paragraph, to derive separate mean-field equations that are specifically tailored to model both types of aforementioned high-lying BEC states.

In order to clarify and organize our discussion of macroscopic quantum self-trapped and superposition states, we present a model that encapsulates the minimal physics necessary to describe their basic properties. In the first case (a), all atoms are condensed into a single Fock state; the expansion coefficients for the $\nu$th excited state are of the self-trapped form ${\bf C}^{\nu}=[0,\ldots,0,1^{\nu}_{N_1},0,\ldots,0].$ In the second case (b), the expansion coefficients mix two configurations with equal weight, {\it i.e.}, ${\bf C}^{\nu}=[0,\ldots,0,({1}/{\sqrt 2})^{\nu}_{N_1},0,\ldots,0,\pm({1}/{\sqrt 2})^{\nu}_{N_2},0,\ldots,0].$ The resulting double-configurational many-body BEC state, which is the simplest multiconfigurational state imaginable, is a superposition involving a macroscopic number of atoms. We choose to consider these extremal model cases because MCBHF theory does not contain the full off-diagonal coupling of $\gamma^\nu$ and $\Gamma^\nu$ to the orbitals $\chi_k$ in the underlying (diagonal) BHF equations. However, in the extreme Fock state limit of (a) and (b), or even in slightly less extreme situations, the off-diagonal couplings contribute nothing or, at most, a small effect. All that remains is the diagonal coupling of the expansion coefficients to the orbitals, which we treat in this section exactly {\it per manum} by deriving specific BHF equations corresponding to each of the above situations. Such equations will be explicated in the following.

Corresponding to case (a), where the many-body state of the BEC is single-configurational, the diagonal Fock space matrix elements of the one- and two-body reduced density matrices are $\gamma^\nu_{kk}=N_k,$ $\Gamma^\nu_{kkkk}=N_k(N_k-1),$ and $\Gamma^\nu_{klkl}=\Gamma^\nu_{kllk}=N_kN_l,$ while in the double-configurational case (b) they are $\gamma^\nu_{kk}=N/2,$ $\Gamma^\nu_{kkkk}=[N_k(N_k-1)+N_l(N_l-1)]/2,$ and $\Gamma^\nu_{klkl}=\Gamma^\nu_{kllk}=N_kN_l.$ Substitution of these matrix elements into the general BHF equations (\ref{GHF}) results in two sets of equations that take into account the diagonal part of the coupling of the model states' expansion coefficients ${\bf C}^\nu$ to the orbitals. The familiar BHF equations for identical bosons \cite{Esry1997c,Cederb04} emerge in case (a), while in case (b), this coupling accounts for the diagonal interaction of the two configurations $(1/\sqrt{2})|N_1,N_2;\{{\bf C}^{\nu};\alpha\}\rangle$ and $\pm(1/\sqrt{2})|N_2,N_1;\{{\bf C}^{\nu};\alpha\}\rangle$ of the superposition state density and results in a new pair of coupled mean-field equations for superposition states. We solve the resulting sets of equations, {\it mutatis mutandis}, using a pseudospectral grid method in quasi-one-dimension \cite{fn7} with standard relaxation methods (see Ref. \cite{Masiello05} for details) where, in addition, the orbitals are constrained to be of the form
\begin{equation}
\begin{split}
\label{WF1}
\chi_1({\bf x})&=\sqrt{\alpha}\chi^L_1({\bf x})+\sqrt{1-\alpha}\chi^R_1({\bf x})\\
\chi_2({\bf x})&=\sqrt{1-\alpha}\chi_2^L({\bf x})+\sqrt{\alpha}\chi_2^R({\bf x})
\end{split}
\end{equation}
with $\chi_k^L$ and $\chi_k^R$ being the left and right half of $\chi_k$ taken with respect to the trap center, and $0\leq\alpha\leq1.$ This functional form allows for the orbital basis to be continuously fractionated from delocalized symmetric and antisymmetric pairs ($\alpha=1/2$) to symmetry-broken orbitals that are approximately localized in the left and right parts of the double-well trapping potential ($\alpha\approx0$ or $\alpha\approx1$). One could imagine this variational parameter as being independent from the occupation number and performing a second variation on $n_k^\nu,$ however, for the purpose of our MCBHF model we further choose $\alpha$ to be consistent with $n_k^\nu;$ {\it i.e.}, we take $\alpha=n^\nu_1/N,$ which equals $N_1/N$ in this extreme Fock state limit. By varying $\alpha$ and then applying the Hylleraas-Undheim theorem \cite{Masiello05} to each of the $N+1$ model MCBHF states $|\Psi^N;\{{\bf C}^\nu;\alpha=n^\nu_k/N\}\rangle,$ we empirically find that this choice implies approximate self-consistency, by construction, for both low-lying and high-lying BEC states. This freedom of fractionation is motivated by the fact that the ground state and excited self-trapped and superposition state energies each find their minima at different values of $\alpha,$ or equivalently, in different orbital bases; the ground state and low-lying excited state energies are minimized in the delocalized (anti)symmetric basis, while the high-lying excited state energies reach their minima in the symmetry-broken left- and right-localized basis. It is well known that these bases are not unitarily equivalent in the mean-field approximation for bosons \cite{Cederb04}.

\begin{figure}
\psfrag{CI and MF energy}[][]{{\LARGE $E^{\nu}$ $(\hbar\omega)$}}
\psfrag{N1/N}[][]{{\LARGE $\alpha=n^\nu_1/N$}}
\psfrag{a}[][]{{\LARGE (a)}}
\psfrag{b}[][]{{\LARGE (b)}}
\rotatebox{0}{\resizebox{!}{6.25cm}{\includegraphics{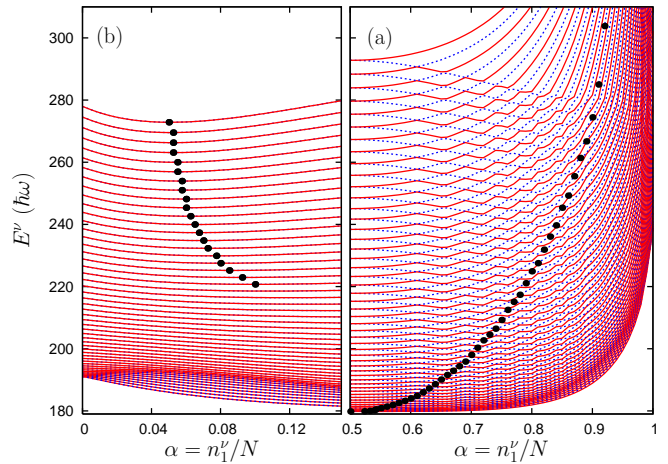}}}
\caption{\label{f1}(Color online) MCBHF energy $E^{\nu}$ as a function of fractionation parameter $\alpha=n^\nu_1/N$ corresponding to macroscopic quantum self-trapped states [case (a)] on the right, and macroscopic quantum superposition states [case (b)] on the left. In both cases, bullets lie at the optimal occupations for each excited state $\nu$ as variationally optimized by the Hylleraas-Undheim theorem. For each self-trapped state $\nu,$ the optimal fractionation $\alpha$ is determined by following curve-crossings. The total number of atoms $N=100$ and the details of trap geometry and appropriately scaled atomic interaction are similar to what is specified in Ref. \cite{Masiello05}. Only the right half of $E^\nu(0.5\leq\alpha\leq1)$ for case (a) and the far left part of $E^\nu(0\leq\alpha\leq0.15)$ for case (b) are displayed.}
\end{figure}

\begin{figure*}
\psfrag{psi}[][]{{\large orbitals $(\beta^{-1/2})$}}
\psfrag{psi1}[][]{{\large trapping potential $(\hbar\omega)$}}
\psfrag{x}[][]{{\large $x$ $(\beta)$}}
\psfrag{density}[][]{{\large density $(\beta^{-1})$}}
\psfrag{ground}[][]{{\large ground\ \ \ }}
\psfrag{CAT}[][]{{\large sp\!\!}}
\psfrag{ST}[][]{{\large s-t}}
\psfrag{a}[][]{{\large (a)}}
\psfrag{b}[][]{{\large (b)}}
\rotatebox{0}{\resizebox{!}{4.5cm}{\includegraphics{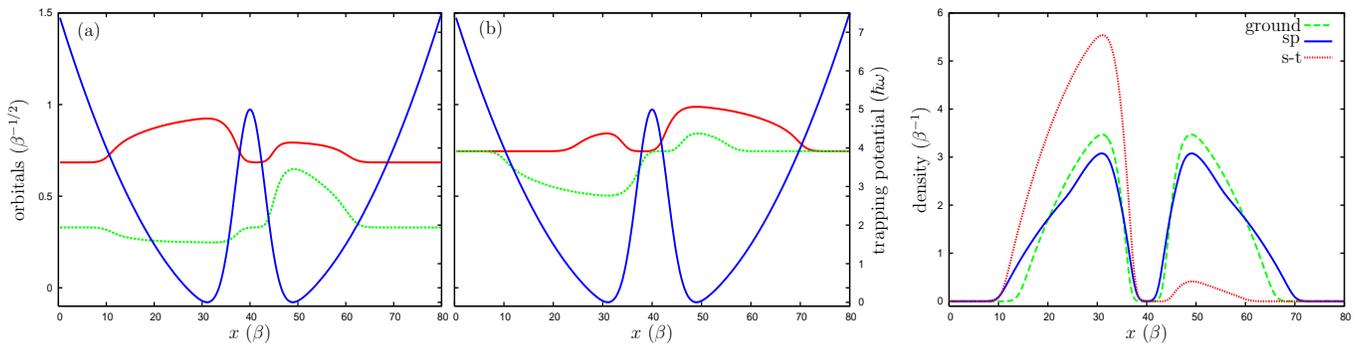}}}
\caption{\label{f2}(Color online) MCBHF orbitals and densities corresponding to the $\nu=90$ self-trapped state, optimized at $\alpha=0.88,$ and the associated superposition state, optimized at $\alpha=0.055.$ The self-trapped orbitals [case (a)] are displayed on the left and the superposition state orbitals [case (b)] are displayed in the center. In both cases, the nodeless orbital is $\chi_1$ and the nodal orbital is $\chi_2.$ They appear in the symmetric double well $V_{\textrm{ext}}$ at their appropriate chemical potentials. The far right panel presents the densities corresponding to each of these states as well as to the condensate ground state.}
\end{figure*}

Full diagonalization of $\hat H$ in the MCBHF states $|\Psi^N;\{{\bf C}^\nu;\alpha=n^\nu_k/N\}\rangle$ corresponding to cases (a) and (b) results in the energy spectra displayed in Fig. \ref{f1} as a function of $\alpha=n_1^\nu/N.$ 
In this, as in all other results shown here, $N=100$ and the scattering length $a$ is chosen so that $aN$ is representative of that value for typical laboratory BECs. The optimal MCBHF occupation numbers for the ground state and each excited self-trapped and superposition state, as variationally optimized by the Hylleraas-Undheim theorem, are labeled with a bullet. Note that, in this approximation, the occupation numbers appearing in the BHF equations may differ from those of the many-body calculation since off-diagonal couplings arise in the latter. For comparison with later MCBSCF results we display in Fig. \ref{f2}, two pairs of BHF orbitals corresponding to the $\nu=90$ self-trapped and superposition states. We point out that each of the high-lying superposition states have energies that lie lower than the corresponding self-trapped state of the same excitation, as can been seen in Fig. \ref{f1}. For example, the $\nu=90$ MCBHF self-trapped state, which is optimized at $\alpha=0.88,$ has an energy of 261.373 $\hbar\omega$ while the equivalent superposition state, which is optimized at $\alpha=0.055,$ has an energy of 256.938 $\hbar\omega.$ Due to the nonlinear dependence of the energy upon the orbitals, this difference is a surprising new result. The spatial densities
\begin{equation}
\rho^\nu({\bf x})=\gamma^\nu({\bf x},{\bf x})=\sum_{kl=1,2}\chi_k^*({\bf x})\gamma^\nu_{kl}\chi_l({\bf x})
\end{equation}
corresponding to these states as well as the MCBHF ground state, which is optimized at $\alpha=0.5,$ are displayed in the rightmost panel of Fig. \ref{f2}.

{\it Full implementation of restricted Fock space model.} Having discussed both self-trapped and superposition states of the gaseous double-well BEC within the framework of our diagonal model, we now study these same excited states within the more general theory that includes the off-diagonal couplings to the many-body densities $\gamma^\nu$ and $\Gamma^\nu$ in the orbital equations [see Eq. (\ref{orb1})], and, in addition, enjoys a variational degree of symmetry breaking.  As before, the solutions of these MCBSCF equations, which are parametrized by $\alpha,$ are obtained with standard relaxation methods. However, in this case, $\alpha$ is not linked to the occupation number; it is an independent variational parameter that controls the degree of spatial symmetry breaking in the underlying orbital basis [see Eq. (\ref{WF1})] and measures the fraction of each orbital's $L^2$-norm on the left- and right-hand sides of $V_{\textrm{ext}}$ taken with respect to the trap center. Rather than applying the Hylleraas-Undheim theorem to the resulting energy eigenvalues of $\hat H$ in order to determine the optimal occupation number for each MCBSCF self-trapped or superposition state $|\Psi^N;\{{\bf C}^\nu;\alpha\}\rangle,$ we instead reach self-consistency in the eigenvalues $n_k^\nu$ of the one-body reduced density matrix $\gamma^\nu({\bf x},{\bf x}')$ separately for each $\nu.$ The Hylleraas-Undheim theorem is then used to variationally optimize the MCBSCF energy $E^\nu$ as a function of fractionation $\alpha.$

\begin{figure*}
\psfrag{psi}[][]{{\large orbitals $(\beta^{-1/2})$}}
\psfrag{psi1}[][]{{\large trapping potential $(\hbar\omega)$}}
\psfrag{x}[][]{{\large $x$ $(\beta)$}}
\psfrag{density}[][]{{\large density $(\beta^{-1})$}}
\psfrag{ground}[][]{{\large ground\ \ \ }}
\psfrag{CAT}[][]{{\large sp\!\!}}
\psfrag{ST}[][]{{\large s-t}}
\psfrag{a}[][]{{\large (a)}}
\psfrag{b}[][]{{\large (b)}}
\rotatebox{0}{\resizebox{!}{4.5cm}{\includegraphics{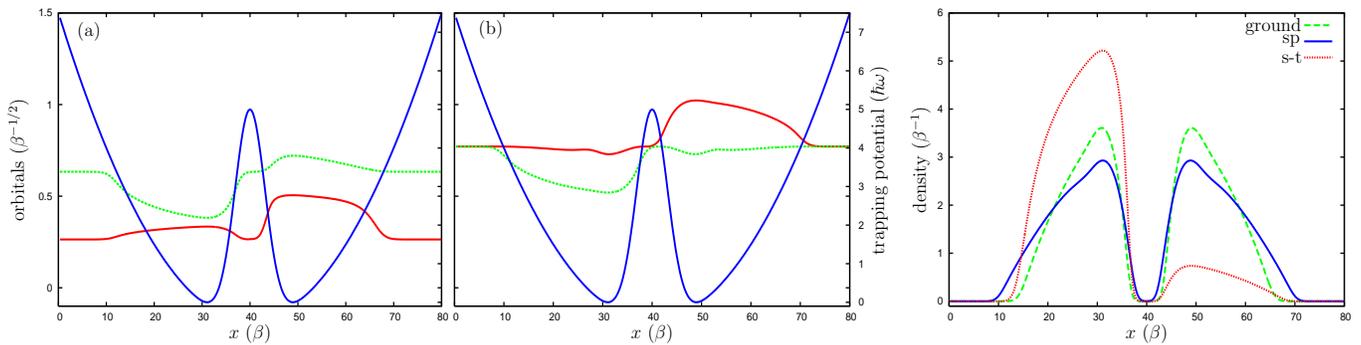}}}
\caption{\label{f3}(Color online) MCBSCF orbitals and densities corresponding to the $\nu=90$ self-trapped state, optimized at $n_1^\nu=11,\alpha=0.06,$ and the associated superposition state, optimized at $n_k^\nu=50,\alpha=0.035.$ The self-trapped orbitals are displayed on the left and the superposition state orbitals are displayed in the center. In both cases, the nodeless orbital is $\chi_1$ and the nodal orbital is $\chi_2.$ They appear in the symmetric double well $V_{\textrm{ext}}$ at their appropriate chemical potentials. The rightmost panel presents the densities corresponding to each of these states as well as to the condensate ground state.}
\end{figure*}

Since the full coupling of $\gamma^\nu$ and $\Gamma^\nu$ to $\chi_k$ is included, there is no need to work within a model as we did previously. The expansion coefficients ${\bf C}^\nu$ may now involve many configurations in both Fock space and orbital sectors and will, by iterating until self-consistency, displace and distort the underlying MCBSCF orbitals through the many-body densities and repulsive coupling constants appearing in Eq. (\ref{orb1}). In fact, self-trapped states owe their stability, in part, to an interference between neighboring configurations. The orbitals corresponding to the $\nu=90$ macroscopic quantum self-trapped state, optimized at $n_1^\nu=11,\alpha=0.06,$ and the associated macroscopic quantum superposition state, optimized at $n_k^\nu=50,\alpha=0.035,$ are displayed in Fig. \ref{f3}. The rightmost panel presents the densities $\rho^\nu$ associated with with each of these states and, in addition, to that of the $\nu=0$ condensate ground state. In contradistinction to the orbitals and densities displayed in Fig. \ref{f2}, it is clear that the MCBSCF based orbitals fill up the entire trap at a given chemical potential. Furthermore, each of the associated densities have noticeably different spatial profiles.

In conclusion, we have developed a novel theoretical method, that is easily extendable to modeling actual experiments \cite{fn7}, for characterizing the high-lying excited states of the gaseous double-well BEC that completely and consistently accounts for the inclusion of atomic correlation and condensate mean-field effects within a sufficiently general Fock space that is flexible enough to describe both macroscopic symmetry breaking and one-body orbital symmetry breaking. Many-body properties of both macroscopic quantum self-trapped and superposition states of the BEC have been illustrated and differences in their energies and densities have been pointed out. Our initial results suggest that Thomas-Fermi theory alone provides a good zeroth order approximation to the ground and high-lying excited state densities of the gaseous double-well BEC, suggesting the possibility of a hitherto unexpected simplification in the large $N$ limit.

The authors gratefully acknowledge financial support from the National Science Foundation through the grant PHY 0140091.

\bibliography{jila,thesis,mal}

\begin{thebibliography}{19}
\bibitem[{fn6()}]{fn6}
\bibinfo{note}{While our theoretical methods are capable of treating asymmetric
  double-well traps, we choose to present only the symmetric double well. Here,
  it is already a challenge to understand the role of symmetry breaking in the
  many-body theory.}
\bibitem[{fn7()}]{fn7}
\bibinfo{note}{Extending Eq. (\ref{orb1}) and its solutions from
  quasi-one-dimension to three-dimensions poses a challenge that may be
  overcome, without difficulty, with increased computing time.}
\end{thebibliography}
\end{document}